\documentclass[english]{appolb}
\usepackage[T1]{fontenc}
\usepackage[latin9]{inputenc}
\usepackage{geometry}
\usepackage{multirow}
\usepackage{amsmath}

\makeatletter

\usepackage{float}
\usepackage{tabu}

\newcolumntype {K}{X[1,c,m]}
\tabucolumn K

\tabulinesep =^3pt
\makeatother

\usepackage{babel}
\begin{document}
\eqsec

\title{Small oscillations of test particles with angular momentum in Reissner-Nordstrom
field}

\author{Valentin D Gladush
\address{vgladush@gmail.com\\
Theoretical physics department, Dnepropetrovsk National University,\\
Gagarina Ave 72, Dnepropetrovsk, 49010, Ukraine}
\and
Alexander I Petrusenko
\address{iftifn@gmail.com\\
Theoretical physics department, Dnepropetrovsk National University,\\
Gagarina Ave 72, Dnepropetrovsk, 49010,
Ukraine}
}

\maketitle

\begin{abstract}
The oscillations of the test particles on circular orbits in field
of super-extremal charged central object are researched. The formulae
for oscillation frequencies are obtained in cases of circular orbits
perturbations caused by small change in energy of the particle or
by the small change in mass of the central object.
\end{abstract}

\section{Introduction}

Radial motion of a charged test particle with orbital angular momentum
in the field of super-extreme charged central body have been considered
in \cite{Gladush-RN+L}. The classification motion types of test particles
have been done and their circular orbits have been described. In this
paper we consider oscillations of a particle caused by a small perturbation
of a circular orbit. The perturbations can be caused by small change
of any of the system parameters $\left\{ M,Q,L,E,m\right\} $. We consider
perturbations that are caused by the change in energy of the particle
and the change in mass of the central object.

The Hamiltonian of the system, the equation of motion and the effective
potential have the form 
\begin{gather}
H=\sqrt{Fm^{2}+F^{2}P_{R}^{2}+F\frac{L^{2}}{R^{2}}},\qquad F=1-\frac{2M}{R}+\frac{Q^{2}}{R^{2}},\\
\left(m\frac{dR}{dt}\right)^{2}=-U_{V}=E^{2}-W_{\varepsilon}^{2},\qquad W_{\varepsilon}^{2}=F\left(m^{2}+\frac{L^{2}}{R^{2}}\right),
\end{gather}
where $M<Q$ is the mass and charge of the central body, $m,L,E$
- the mass, the orbital angular momentum and energy of the test particle.
Minimum of the effective potential $W_{\varepsilon}^{2}$ determines
the circular orbit $\left(R=R_{extr}\right)$ of a particle.

\section{Perturbations caused by changes in the particle energy}

In this case, small deviations of the particle\textquoteright{}s world
line can be reduced to a canonical transformation $R=R_{extr}+r,\quad P_{R}=P_{r}$.
Expanding the new Hamiltonian near an equilibrium position, and taking
into account that the first derivatives and mixed derivatives equal
zero, we obtain
\begin{gather}
H(P_{R},R)\rightarrow H(P_{r},r)=E_{min}+\frac{1}{2}H_{PP}P_{r}^{2}+\frac{1}{2}H_{rr}r^{2},\\
H_{PP}=\left(\frac{\partial^{2}H}{\partial P_{r}^{2}}\right)_{r=0,P_{r}=0}=\left(\frac{F^{2}}{E_{min}}\right)_{R=R_{extr}},\\
H_{rr}=\left(\frac{\partial^{2}H}{\partial r^{2}}\right)_{r=0,P_{r}=0}=\left(\frac{\frac{d^{2}W_{\varepsilon}}{dR^{2}}}{2E_{min}}\right)_{R=R_{extr}}.
\end{gather}
Comparing this Hamiltonian with a classical oscillator Hamiltonian,
we obtain the expression for the oscillation frequency
\begin{gather}
H_{PP}=\frac{1}{m_{eff}},\qquad k=H_{rr},\qquad\omega=\sqrt{\frac{k}{m_{eff}}}=\sqrt{H_{rr}H_{PP}},\\
\omega=\frac{1}{\sqrt{2}}\left(\frac{F}{E_{min}}\sqrt{\frac{d^{2}W_{\varepsilon}}{dR^{2}}}\right)_{R=R_{extr}}.\label{eq:1.7}
\end{gather}
The formula for $\omega$ is not given in terms of $E_{min},R_{extr}$
in order to reduce the paper length.

\section{Perturbations caused by changes in mass of the central object}

The dynamics of test particles with orbital angular momentum can also
be studied using the mass potential \cite{Glad-Mass-pot}. The velocity
potential has the form
\begin{gather}
U_{V}\left(M,Q,E,m,L\right)=-E^{2}+\left(1-\frac{2M}{R}+\frac{Q^{2}}{R^{2}}\right)\left(m^{2}+\frac{L^{2}}{R^{2}}\right).
\end{gather}
Definition of the mass potential is as follows $U_{V}\left(U_{M},Q,E,m,L\right)=0$.
Hence, the mass potential is 
\begin{gather}
U_{M}=\frac{1}{2R}\left[\left(R^{2}+Q^{2}\right)-\frac{R^{2}E^{2}}{\left(m^{2}+\frac{L^{2}}{R^{2}}\right)}\right].
\end{gather}
Perturbations of the mass of the central object can be reduced to
the transformation $U_{M}=M+\mu$. Velocity and acceleration of a
particle in terms of the mass potential are
\begin{gather}
\left(\frac{dR}{ds}\right)^{2}=\frac{2}{R}\left(M-U_{M}\right),\label{eq:13}\\
\frac{d^{2}R}{ds^{2}}=-\frac{1}{R^{2}}\left(M-U_{M}+R\frac{dU_{M}}{dR}\right).\label{eq:14}
\end{gather}
Let us expand the mass potential in the vicinity of  $R_{min,}\quad R=R_{min}+r,$
\begin{gather}
U_{M}\left(R\right)\rightarrow U_{M}\left(r\right)=U_{M}\left(0\right)+r\left(\frac{dU_{M}}{dr}\right)_{r=0}+\frac{r^{2}}{2}\left(\frac{d^{2}U_{M}}{dr^{2}}\right)_{r=0},\label{eq:15}\\
U_{M0}^{''}=\left(\frac{d^{2}U_{M}}{dr^{2}}\right)_{r=0}=\left(\frac{d^{2}U_{M}}{dR^{2}}\right)_{R=R_{min}}=\frac{1}{R}\left[\frac{Q^{2}}{R^{2}}-\frac{L^{2}}{R^{2}}\frac{E^{2}}{\left(m^{2}+\frac{L^{2}}{R^{2}}\right)^{2}}\left(\frac{4\frac{L^{2}}{R^{2}}}{m^{2}+\frac{L^{2}}{R^{2}}}-1\right)\right].\label{eq:17}
\end{gather}
Substituting this expansion into formulas for velocity and acceleration
of the particle and neglecting all second and higher orders infinitesimals
we obtain the equation of the oscillator and the oscillation frequency
as follows
\begin{gather}
\frac{d^{2}r^{'}}{ds^{2}}+\frac{U_{M0}^{''}}{R_{min}}r^{'}=0,\qquad r^{'}=r+\frac{\mu}{U_{M0}^{''}R_{min}},\\
\tilde{\omega}=\sqrt{\frac{U_{M0}^{''}}{R_{min}}}=\frac{1}{R_{min}}\sqrt{\frac{Q^{2}}{R_{min}^{2}}-\frac{L^{2}}{R_{min}^{2}}\frac{E_{min}^{2}}{\left(m^{2}+\frac{L^{2}}{R_{min}^{2}}\right)^{2}}\left(\frac{4\frac{L^{2}}{R_{min}^{2}}}{m^{2}+\frac{L^{2}}{R_{min}^{2}}}-1\right)}.\label{eq:19}
\end{gather}
Coordinate transformation  $r\rightarrow r^{'}$ is interpreted as
the shift of the center of particle oscillations caused by the change
in mass of the central object.

Note that the frequency of small oscillations of the particles $\omega$
(see (\ref{eq:1.7})) is measured relative to the time an infinitely
distant observer $T$, whereas the frequency $\tilde{\omega}$ (see
(\ref{eq:19})) - relative to the proper time. The oscillation frequency
$\varpi_{\infty}$ of a source, measured by a distant observer, and
the oscillation frequency $\tilde{\varpi}$ of the same source relative
to the proper time are connected by the relation
\[
\varpi_{\infty}=\tilde{\varpi}\sqrt{g_{00}}=const.
\]
Therefore, in the case of small oscillations of the particles around
the equilibrium position, we have the following relation between frequencies
(\ref{eq:1.7}) and (\ref{eq:19})
\[
\omega=\tilde{\omega}\sqrt{F_{extr}}=\tilde{\omega}\sqrt{1-\frac{2M}{R_{extr}}+\frac{Q^{2}}{R_{extr}^{2}}}.
\]

\section{Tables of the oscillation frequencies for the substantially different
types of system behavior}

Types of system behavior can be divided by type of roots of circular
orbit equation; it has been done in \cite{Gladush-RN+L}. There have
also been highlighted regions of parameters for each type of behavior
and calculated circular orbits. In this work we obtained the values
of oscillation frequencies for the two types of perturbations for
each of the parameters sets with significantly different behavior
of the system. 

Following \cite{Gladush-RN+L} we use dimensionless quantities 
\begin{gather*}
\varepsilon=\frac{E}{mc^{2}},\sigma=\frac{\left|Q\right|m\sqrt{k}}{c\left|L\right|},\mu=\frac{Mmk}{c\left|L\right|},z=\frac{Rmc}{\left|L\right|},\Omega=\frac{\left|L\right|}{mc^{2}}\omega,\tilde{\Omega}=\frac{\left|L\right|}{mc^{2}}\tilde{\omega}.
\end{gather*}
Thus the frequency is represented in units of $\left|L\right|/mc^{2}$.
We can estimate this value; for example, for a central body with mass
of the sun this value is measured in kilohertz. The $\Omega$ is the
frequency for small deviations of $\varepsilon$, and the $\tilde{\Omega}$ is
the frequency for small deviations of $\mu$.

Following tables contain computed oscillation frequencies $\Omega,\tilde{\Omega}$
for the data $\left(z_{min},\varepsilon_{min}\right)$ obtained in
\cite{Gladush-RN+L} for given system parameters $\left(\sigma,\mu\right)$. Complex values of oscillation frequencies $\Omega,\tilde{\Omega}$ correspond to the non-stable circular orbits.

\begin{table}[H]
\caption{Region $D_{1}^{(3)}$}
\begin{tabu}{|c|K|K|K|K|}
\hline 
\multicolumn{5}{|c|}{$D_{1}^{(3)}$, system parameters  $\sigma^{2}=0.1,\quad\mu^{2}=0.088$}\tabularnewline
\hline 
\hline 
  & Radius of the orbit $z_{m}$ & Energy $\varepsilon_{min}$  & Frequency of small oscillations $\Omega$ & Frequency of small oscillations $\tilde{\Omega}$\tabularnewline
\hline 
$z_{m1}$ & $2.6839$ & $0.9502$ & $0.08354$ & $0.09382$\tabularnewline
\hline 
$z_{m2}$ & $0.40706$ & $1.0134$ & $0.5133$ & $1.3434$\tabularnewline
\hline 
$z_{m3}$ & $0.61713$ & $0.61713$ & $0.5854\, i$ & $0.62998\, i$\tabularnewline
\hline 
\end{tabu}
\end{table}

\begin{table}[H]
\caption{Region $D_{2}^{(3)}$}
\begin{tabu}{|c|K|K|K|K|}
\hline 
\multicolumn{5}{|c|}{$D_{2}^{(3)}$, system parameters $\sigma^{2}=0.16,\quad\mu^{2}=0.13607$}\tabularnewline
\hline 
\hline 
  & Radius of the orbit $z_{m}$ & Energy $\varepsilon_{min}$  & Frequency of small oscillations $\Omega$ & Frequency of small oscillations $\tilde{\Omega}$\tabularnewline
\hline 
$z_{m1}$ & $1.6325$ & $0.91449$ & $0.1012$ & $0.1298$\tabularnewline
\hline 
$z_{m2}$ & $0.55539$ & $0.89857$ & $0.3681$ & $0.8435$\tabularnewline
\hline 
$z_{m3}$ & $0.95681$ & $0.91902$ & $0.1551\, I$ & $0.2441\, I$\tabularnewline
\hline 
\end{tabu}
\end{table}

\begin{table}[H]
\caption{Region $\Sigma_{+}^{(2)}$}
\begin{tabu}{|c|K|K|K|K|}
\hline 
\multicolumn{5}{|c|}{$\Sigma_{+}^{(2)}$, system parameters $\sigma^{2}=0.16,\quad\mu^{2}=0.13989$}\tabularnewline
\hline 
\hline 
  & Radius of the orbit $z_{m}$ & Energy $\varepsilon_{min}$  & Frequency of small oscillations $\Omega$ & Frequency of small oscillations $\tilde{\Omega}$\tabularnewline
\hline 
$z_{m1}$ & $0.50947$ & $0.84792$ & 0.4534 & $1.1785$\tabularnewline
\hline 
$z_{m2}=z_{m3}$ & $1.296$ & $0.90913$ & $0$ & $0.002125\, I$\tabularnewline
\hline 
\end{tabu}
\end{table}

\begin{table}[H]
\caption{Region $\Sigma_{-}^{(2)}$}
\begin{tabu}{|c|K|K|K|K|}
\hline 
\multicolumn{5}{|c|}{$\Sigma_{-}^{(2)}$, system parameters $\sigma^{2}=0.1,\quad\mu^{2}=0.085442$}\tabularnewline
\hline 
\hline 
  & Radius of the orbit $z_{m}$ & Energy $\varepsilon_{min}$  & Frequency of small oscillations $\Omega$ & Frequency of small oscillations $\tilde{\Omega}$\tabularnewline
\hline 
$z_{m1}$ & $2.7691$ & $0.95211$ & $0.08108$ & $0.09054$\tabularnewline
\hline 
$z_{m2}=z_{m3}$ & $0.4971$ & $1.0742$ & $0.005439\, I$ & $0.009074\, I$\tabularnewline
\hline 
\end{tabu}
\end{table}

\begin{table}[H]
\caption{Region $\Gamma^{(1)}$}
\begin{tabu}{|c|K|K|K|K|}
\hline 
\multicolumn{5}{|c|}{$\Gamma^{(1)}$, system parameters $\sigma^{2}=0.2,\quad\mu^{2}=0.16$}\tabularnewline
\hline 
\hline 
  & Radius of the orbit $z_{m}$ & Energy $\varepsilon_{min}$  & Frequency of small oscillations $\Omega$ & Frequency of small oscillations $\tilde{\Omega}$\tabularnewline
\hline 
$z_{m1}=z_{m2}=z_{m3}$ & $1$ & $2/\sqrt{5}$ & $0.\, I$ & $0$\tabularnewline
\hline 
\end{tabu}
\end{table}

\begin{table}[H]
\caption{Region  $D_{1}^{(1)}$}
\begin{tabu}{|c|K|K|K|K|}
\hline 
\multicolumn{5}{|c|}{$D_{1}^{(1)}$, system parameters $\sigma^{2}=0.4,\quad\mu^{2}=0.2$}\tabularnewline
\hline 
\hline 
  & Radius of the orbit $z_{m}$ & Energy $\varepsilon_{min}$  & Frequency of small oscillations $\Omega$ & Frequency of small oscillations $\tilde{\Omega}$\tabularnewline
\hline 
$z_{m1}$ & $2.1114$ & $0.90309$ & $0.1351$ & $0.1657$\tabularnewline
\hline 
\end{tabu}
\end{table}

\begin{table}[H]
\caption{Region $D_{2}^{(1)}$}
\begin{tabu}{|c|K|K|K|K|}
\hline 
\multicolumn{5}{|c|}{$D_{2}^{(1)}$, system parameters $\sigma^{2}=0.1625,\quad\mu^{2}=0.15177$}\tabularnewline
\hline 
\hline 
  & Radius of the orbit $z_{m}$ & Energy $\varepsilon_{min}$  & Frequency of small oscillations $\Omega$ & Frequency of small oscillations $\tilde{\Omega}$\tabularnewline
\hline 
$z_{m1}$ & $0.44716$ & $0.64921$ & $0.4874$ & $1.8389$\tabularnewline
\hline 
\end{tabu}
\end{table}

\begin{table}[H]
\caption{Region $D_{3}^{(1)}$}
\begin{tabu}{|c|K|K|K|K|}
\hline 
\multicolumn{5}{|c|}{$D_{2}^{(1)}$, system parameters $\sigma^{2}=0.125,\quad\mu^{2}=0.09$}\tabularnewline
\hline 
\hline 
  & Radius of the orbit $z_{m}$ & Energy $\varepsilon_{min}$  & Frequency of small oscillations $\Omega$ & Frequency of small oscillations $\tilde{\Omega}$\tabularnewline
\hline 
$z_{m1}$ & $2.7781$ & $0.95074$ & $0.08264$ & $0.09238$\tabularnewline
\hline 
\end{tabu}
\end{table}


\begin{thebibliography}{References}
\bibitem{Gladush-RN+L} Gladush V.D., Kulikov D.A. Classification
for the radial component of particles motion in the field of a super-extremely
charged object. \textbf{arXiv:1110.3179} \textbf{{[}gr-qc{]}}.

\bibitem{Glad-Mass-pot} Gladush V.D., Galadgyi M.V. Some peculiarities
of motion of neutral and charged test particles in the field of a
spherically symmetric charged object in general relativity. \textbf{Gen
Relativ Gravit. 43}, 1347\textendash{}1363 (2011).\end{thebibliography}
\end{document}